\documentclass{article}
\usepackage{spconf,amsmath,graphicx}
\usepackage{xcolor}
\usepackage{multirow}
\usepackage{cite}

\title{Multi-stage Speaker Extraction with Utterance and Frame-Level Reference Signals}
\name{Meng Ge$^{1,2}$, Chenglin Xu$^{4,*}$, Longbiao Wang$^{1, *}$, Eng Siong Chng$^2$, Jianwu Dang$^{1, 3}$, Haizhou Li$^{4,5}$ 
\thanks{This research is supported by the National Research Foundation, Singapore under its AI Singapore Programme (AISG Award No: AISG-100E-2018-006); Human-Robot Interaction Phase 1 (Grant No. 192 25 00054), National Research Foundation (NRF) Singapore under the National Robotics Programme; National Natural Science Foundation (61771333), Tianjin Municipal Science and Technology Project (18ZXZNGX00330).
This research is also funded by the Deutsche Forschungsgemeinschaft (DFG, German
Research Foundation) under Germany's Excellence Strategy (University Allowance, EXC 2077,
University of Bremen).
$^*$ Corresponding author.}}
\address{
  $^1$ Tianjin Key Laboratory of Cognitive Computing and Application,\\ College of Intelligence and Computing, Tianjin University, Tianjin, China\\
  $^2$ School of Computer Science and Engineering, Nanyang Technological University, Singapore\\
  $^3$ Japan Advanced Institute of Science and Technology, Ishikawa, Japan\\
  $^4$ Department of Electrical and Computer Engineering, National University of Singapore, Singapore \\
  $^5$ Machine Listening Lab, University of Bremen, Germany}
  
\begin{document}
%
\maketitle
\begin{abstract}
Speaker extraction requires a sample speech from the target speaker as the reference. However, enrolling a speaker with a long speech is not practical. We propose a speaker extraction technique, that performs in multiple stages to take full advantage of short reference speech sample. The extracted speech in early stages is used as the reference speech for late stages.  For the first time, we use frame-level sequential speech embedding as the reference for target speaker. This is a departure from the traditional utterance-based speaker embedding reference. In addition, a signal fusion scheme is proposed to combine the decoded signals in multiple scales with automatically learned weights.
Experiments on WSJ0-2mix and its noisy versions (WHAM! and WHAMR!) show that SpEx++ consistently outperforms other state-of-the-art baselines.
\end{abstract}
\section{Introduction}
\label{sec:intro}
Real-world speech communication usually takes place in complex auditory scenes in the presence of multiple speakers. Blind speech separation is one of the solutions to isolate one source from others, such as DPCL  \cite{hershey2016deep,isik2016single,wang2018alternative}, PIT \cite{yu2017permutation,kolbaek2017multitalker,xu2018single}, and TasNet \cite{luo2018real,luo2019conv,luo2020dual}. However, it requires that the number of sources is known in advance, and assumes the permutation of source labels is unchanged during training, which greatly limits its scope of applications. 

Unlike blind speech separation, speaker extraction only extracts the target speech from a mixture speech given a reference utterance of the target speaker \cite{vzmolikova2019speakerbeam,wang2018voicefilter,xu2019optimization,xu2019time,spex2020,spex_plus2020}. Thus, it naturally avoids the problems of arbitrary source permutation and unknown number of sources. Previous studies \cite{spex2020,spex_plus2020} show that longer duration of a reference speech in speaker extraction always leads to better performance. However, real applications usually only permit a short reference utterance, e.g., a wakeup word for mobile device. This prompts us to study how to reduce the required duration of a reference utterance.

\begin{figure*}[t]
	\centering
	\includegraphics[width=0.66\linewidth]{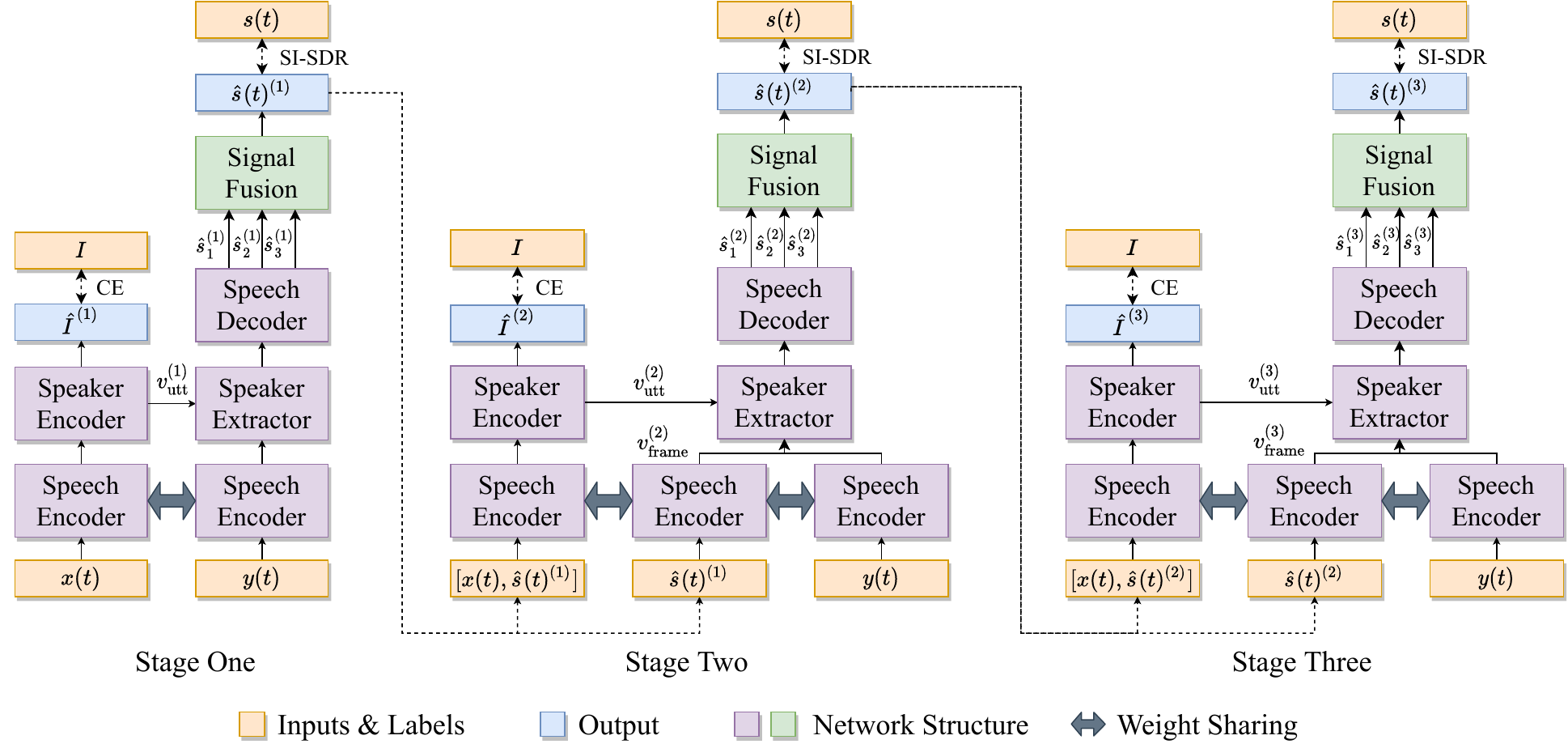}
	\caption{The diagram of the SpEx++ system. Each stage consists of speech encoders, speaker encoder, speaker extractor, speech decoder and signal fusion. The $y(t)$, $x(t)$, $s(t)$, and $I$ are the mixture speech, reference speech, clean speech and true speaker label, respectively. The $\hat{s}(t)$, $\hat{I}$, $v_{\text{utt}}$ and $v_{\text{frame}}$ represent the extracted target voice, the predicted probability of target speaker, utterance-level speaker embedding and newborn frame-level speaker embedding in different stages, respectively.}
	\label{fig:multi-stage_framework}
	\vspace{-10pt}
\end{figure*}

Psychoacoustic studies \cite{Kuhl1991magneteffect} suggest brain circuits create perceptual attractors, or magnets, that warp the stimulus space such that it draws the sound that is closest to it. Human auditory attention uses the current attended acoustic stimulus to reinforce the attractor \cite{bronkhorst2015cocktail} in a continuous auditory process. In speaker extraction, the reference speech that is encoded as a speaker embedding can be seen as such a perceptual attractor. Motivated by the psychoacoustic studies, we propose a speaker extraction architecture, that reinforces the attractor for target speaker in multiple stages. 

Specifically, we take advantage of the multi-stage architecture to reuse the extracted target voice from previous stages to strengthen the attractor. The extracted target speech is combined with the original reference utterance to strengthen utterance-level speaker embedding. At the same time, the extracted speech is utilized to capture the frame-level speaker embedding which is aligned with the mixture speech frame-by-frame. In this way, the extracted speech provides a second reference signal. Note that the first reference signal is an utterance-level speaker embedding, which represents the overall characteristics of target speaker; the second reference signal is a sequential reference that describes the temporal dynamics of what exactly the target speaker speaks. Such information is hard to be captured directly in one single-stage. In addition, we propose a signal fusion strategy to combine the decoded signals in multiple scales with automatically learned weights to produce a higher-quality target speech. 



\section{SpEx++ Architecture}
\label{sec:format}

SpEx++ is a multi-stage SpEx+~\cite{spex_plus2020} pipeline. Without loss of generality, we study a 3-stage architecture in this work, as shown in Fig. \ref{fig:multi-stage_framework}. However, SpEx++ has an entirely different target speaker referencing mechanism from SpEx+. Let's begin with a  brief review of SpEx+ system. We then discuss a signal fusion strategy and a multi-stage system.

\subsection{Review of SpEx+ system}
\label{sec:pagestyle}

SpEx+ system is a complete time-domain speaker extraction solution, that accepts a time-domain mixture speech and reference speech as inputs to extract the target speech. SpEx+ mainly consists of four modules: twin speech encoder, speaker encoder, speaker extractor, and speech decoder. The twin speech encoder shares the network structure and parameters, projecting the mixture speech and the reference speech of the target speaker into a common latent space. The speaker encoder produces the discriminative utterance-level speaker embedding from the reference speech. The speaker extractor collects the utterance-level speaker embedding from speaker encoder and transformed features from twin speech encoder to estimate masks in various scales, and then goes through the speech encoder to extract the target signals in various scales. 

SpEx+ network is trained with multi-task learning objective, that measures both the target speaker prediction and signal reconstruction. The former minimizes the Cross-Entropy (CE) loss between predicted target speaker and actual speaker label. The latter minimizes reconstruction error (multi-scale SI-SDR) between the extracted signals in different scales and the clean target speech.

SpEx+ processes speech in multiple scales. However, at run-time inference, the extracted target signal with smallest scale (highest temporal resolution) is chosen as the final output~\cite{spex_plus2020}. As a result, the decoded signals from other scales don't contribute to target speech generation. In addition, the weights to balance the multi-scale SI-SDR loss between the extracted signals in different scales and the clean target speech are tuned on the development set, which adds complexity to the training process.

\subsection{Fusion of multi-scale signals}
\label{sec:signal_fusion}

{To take advantage of the complementary decoded signals from multiple scales during training and inference,}
we propose a signal fusion strategy before the loss calculation. First, we combine multi-scale decoded signals with automatically learned weights, and then calculate the reconstruction error (SI-SDR) between the fused signal and clean signal. 

{Let $\hat{s}_1^{(k)}$, $\hat{s}_2^{(k)}$, and $\hat{s}_3^{(k)}$ denote the multi-scale decoded signals in stage $k$, we can use several learnable weights ($w_1$, $w_2$, and $w_3$) to obtain the fused signal as follow:
\begin{equation}
    \label{eq:signal_fusion}
    \hat{s}^{(k)} = w_1 * \hat{s}_1^{(k)} + w_2 * \hat{s}_2^{(k)} + w_3 * \hat{s}_3^{(k)}, \quad k=1,2,3
    \vspace{-8pt}
\end{equation}
}

It was reported~\cite{spex2020} that the decoded signal with the highest temporal resolution achieves the best performance, and the weights to balance the multi-scale SI-SDR loss are tuned to be $0.8$, $0.1$, $0.1$ for high, middle, and low temporal resolutions~\cite{spex2020,spex_plus2020}. We follow the same weights to combine the signals of 3 temporal resolutions first, then calculate the SI-SDR between the fused signal and the clean signal as,
\begin{gather}
    \label{eq:sisdr}
    \mathcal{L}_{\text{SI-SDR}}^{(k)} = -\rho(\hat{s}^{(k)}, s), \\
    \rho(\hat{s}, s) = 20 \log_{10}\frac{||(\hat{s}^T s / s^T s) \cdot s||}{||(\hat{s}^T s / s^T s) \cdot s - \hat{s}||}
\end{gather}
where $\hat{s}$ and $s$ are the estimated signal and the target clean signal, respectively.

\subsection{Multi-stage speaker extraction}
\label{sec:pagestyle}

The idea of multi-stage architecture is to pipeline  several single-stage speaker extraction modules sequentially such that a later module operates on the extracted target speech of an earlier one. The effect of such composition is an incremental refinement of the reference speech from the target speaker. As shown in Fig. 1, each stage takes the extracted target voice from the previous stage as an extra input to refine the reference speech of the target speaker. We implement two mechanisms in SpEx++ to do so.

One  is to combine the extracted target voice $\hat{s}^{(k-1)}(t)$ with the original, short reference utterance $x(t)$ to extract the utterance-level speaker embedding $v_{\text{utt}}^{(k)}$. Such speaker embedding is expected to represent characteristics of the target speaker, that is referred to as the first reference signal,
\begin{equation}
    v_{\text{utt}}^{(k)} = \text{Enc}_{\text{speaker}}(\text{Enc}_{\text{speech}}(\text{Concat} (x, \hat{s}^{(k-1)}))),  
\end{equation}
where $k=2,3$, $\text{Enc}_{\text{speech}}(\cdot)$ and $\text{Enc}_{\text{speaker}}(\cdot)$ represent the speech encoder and speaker encoder, respectively.

{Another is to pass the extracted speech $s^{(k-1)}(t)$ through the speech encoder to obtain a frame-level speech embedding. As  $s^{(k-1)}(t)$ is obtained from the mixture speech $y(t)$, and aligned with the mixture speech frame-by-frame. Such frame-level embedding is different from the utterance-level speaker embedding, but provides a strong signal about the target speaker and his/her speech content. For the first time, we introduce the frame-level speech embedding $v_{\text{frame}}^{(k)} (t)$ as a second reference signal in speaker extraction study,}
\begin{equation}
    v_{\text{frame}}^{(k)} (t) = \text{Enc}_{\text{speech}}(\hat{s}^{(k-1)}(t)), \quad k=2,3
\end{equation}

Now we formulate the way to use two reference signals for speaker extraction,
\begin{gather}
    M_{i}^{(k)} = \text{Ext}(v_{\text{utt}}^{(k)}, \text{Concat}(v_{\text{frame}}^{(k)}, \text{Enc}_{\text{speech}}(y))), \\
    \hat{s}_i^{(k)} = \text{Dec}(M_{i}^{(k)} \otimes \text{Enc}_{\text{speech}}(y))
\end{gather}
where $i=1,2,3$ represents the three different scales, and $\otimes$ is an operation for element-wise multiplication. The $\text{Ext}(\cdot)$ and $\text{Dec}(\cdot)$ represent the speaker extractor and speech decoder, respectively. Finally, the reconstruction error can be calculated by applying the operations in Section \ref{sec:signal_fusion}.

\renewcommand{\arraystretch}{1.5}
\begin{table}[tp]
	
	\centering
	\fontsize{6.2}{6}\selectfont
	\caption{A comparative study on the WSJ0-2mix dataset under open condition. \textit{Given Ref.} denotes the duration of given test reference speech in second (s). \textit{Target Ref.} shows the use of reference signals. \textit{Utt} denotes utterance-level speaker embedding $v_{\text{utt}}^{(k)}$, \textit{Fm} denotes frame-level speech embedding  $v_{\text{frame}}^{(k)}$. \textit{\#stage} denotes the number of SpEx decoding stages. } 
	\begin{tabular}{|c|c|c|c|c|c|c|c|}
		\hline
		\multirow{2}{*}{\shortstack{Given\\Ref.}} &\multirow{2}{*}{Methods} & \multirow{2}{*}{\shortstack{Signal\\Fusion}}  & \multirow{2}{*}{\#stages} & \multirow{2}{*}{\shortstack{Target\\Ref.}} &\multirow{2}{*}{\shortstack{SDRi\\(dB)}} &\multirow{2}{*}{\shortstack{SI-SDRi\\(dB)}} &\multirow{2}{*}{PESQ} \cr
		 & &  &  &  &  & & \cr
		\hline
		\hline
		\multirow{6}{*}{\shortstack{7.3s\\(avg)}} &SpEx+ \cite{spex_plus2020} &$\times$ & 1 & Utt &17.2 &16.9 &3.43 \\
		&SpEx++ &$\surd$ & 1 & Utt &17.5 &17.2 &3.46\\ 
		&SpEx++  &$\surd$ & 2 & Utt  &17.7 &17.3 &3.47 \\
		&SpEx++  &$\surd$ & 2 & Fm &18.2 &17.8 &3.51 \\
		&SpEx++  &$\surd$ & 2 & Utt+Fm  &18.3 &17.9 &3.52 \\
		&SpEx++  &$\surd$ & 3 & Utt+Fm &\textbf{18.4} &\textbf{18.0} &\textbf{3.53}\\
		\hline
		\hline
		\multirow{4}{*}{2s} &SpEx+ \cite{spex_plus2020} &$\times$ & 1 & Utt &16.6 &16.2 &3.37 \\
		&SpEx++ &$\surd$ & 1 & Utt &16.9 &16.4 &3.40 \\
		&SpEx++  &$\surd$ & 2 & Utt+Fm  &\textbf{17.6}  &17.1  &\textbf{3.46}  \\
		&SpEx++  &$\surd$ & 3 & Utt+Fm &\textbf{17.6} &\textbf{17.2} &\textbf{3.46}
		\cr
		\hline
	\end{tabular} \label{tbl:cmp_system_wsj0_2mix}
	\vspace{-10pt}
\end{table}

\section{Experiments and Discussion}
\label{sec:typestyle}


We evaluated our system on the noise-free two-speaker mixture database WSJ0-2mix. The database was derived from the WSJ0 corpus at sampling rate of 8kHz. The simulated database contained 101 speakers and was divided into three sets: training set (20,000 utterances), development set (5,000 utterances), and test set (3,000 utterances). Specifically, the utterances from two speakers in WSJ0 ``si\_tr\_s" corpus were randomly selected to generate the training and development set at various SNR between 0dB and 5dB. Similarly, the test set was generated by randomly mixing the utterances from two speakers in WSJ0 ``si\_dt\_05" and ``si\_et\_05" set. Since the speakers in test set were unseen during training, the test set was considered as open condition evaluation.

The noisy versions of WSJ0-2mix, called WHAM! \cite{Wichern2019WHAM} and WHAMR! \cite{Maciejewski2020WHAMR}, were also used to verify the robustness of our system. WHAM! paired each two-speaker mixture in WSJ0-2mix with a non-speech ambient noise sample, recorded in real cases such as coffee shops, restaurants, and bars. WHAMR! extended WHAM! by introducing reverberation to the speech sources in addition to the existing noise.

Unlike in blind speech separation, speaker extraction technique requires a reference speech of target speaker as input. For each mixture speech in WSJ0-2mix, the speakers in mixed speech acted as the target speaker in turn, and the corresponding reference speech was randomly selected from original WSJ0 corpus. The reference speeches in WHAM! and WHAMR! were the same as that in WSJ0-2mix.

\subsection{Experimental setup}
We trained all systems for 100 epochs on the mixture segments with 4-second and their corresponding reference utterances as in~\cite{spex_plus2020}, which has an average of 7.3 seconds.  The learning rate was initialized to $1e^{-3}$ and decays by 0.5 if the accuracy of validation set was not improved in 2 consecutive epochs. Early stopping was applied if no best model is found in the validation set for 6 consecutive epochs. Adam was used as the optimizer. The network structure follows SpEx+ \cite{spex_plus2020}. 

The speech encoders at each stage shared the network structure and weights. The filter lengths of convolutions in speech encoder and decoder were 2.5ms, 10ms, and 20ms for the 3 time-scales with speech of 8kHz sampling rate, respectively. The speaker extractor repeated a stack of 8 temporal convolution network (TCN) blocks for 4 times. {We included 3 ResNet blocks in speaker encoder, with 256, 256, and 512 filters respectively.} The utterance-level speaker embedding dimension was set to 256 in practice. For the signal fusion configuration, the weights are initialized as $w_1=0.8$, $w_2=0.1$, $w_3=0.1$, and then learned automatically during training.

\renewcommand{\arraystretch}{1.5}
\begin{table}[tp]
	
	\centering
	\fontsize{6.5}{5.5}\selectfont
	\caption{SDRi (dB) and SI-SDRi (dB) in a comparative study on the WHAM! dataset under the open condition. For blind speech separation (BSS) task, we report the results evaluated on the oracle-selected streams. For speaker extraction (SE) task, we report the results on the extracted target streams.} 
	\begin{tabular}{|c|c|c|c|c|}
		\hline
		Task &Methods &Given Ref. &SDRi &SI-SDRi\cr
		\hline
		\hline
		\multirow{2}{*}{BSS} &BLSTM-TasNet \cite{Wichern2019WHAM} &- &- &9.8 \\
		&chimera++ \cite{Wichern2019WHAM} &- &- &9.9 \\\hline
		\multirow{4}{*}{SE} &SpEx \cite{spex2020} &7.3s (avg) &13.0 &12.2\\
		&SpEx+ \cite{spex_plus2020} &7.3s (avg) &13.6 &13.1\\\cline{2-5}
		&\multirow{2}{*}{SpEx++ (2-stages)} &7.3s (avg) &14.4 &14.0 \\ 
		& &60s &\textbf{14.7} &\textbf{14.3}  \cr
		\hline
	\end{tabular} \label{tbl:cmp_wham}
	\vspace{-10pt}
\end{table}

\renewcommand{\arraystretch}{1.7}
\begin{table}[tp]
	
	\centering
	\fontsize{7}{5}\selectfont
	\caption{SDRi (dB) and SI-SDRi (dB) in a comparative study on the WHAMR! dataset under the open condition.} 
	\begin{tabular}{|c|c|c|c|c|c|}
		\hline
		\multirow{2}{*}{Cond.} &\multirow{2}{*}{Task} &\multirow{2}{*}{Methods} & \multirow{2}{*}{\shortstack{Given\\Ref.}} &\multirow{2}{*}{\shortstack{SDRi\\(dB)}} &\multirow{2}{*}{\shortstack{SI-SDRi\\(dB)}} \cr
		& & & & & \cr
		\hline
		\hline
		\multirow{7}{*}{Noise}&\multirow{3}{*}{BSS} &Conv-TasNet \cite{Maciejewski2020WHAMR} &-  &- &11.5\\
		& &BLSTM-TasNet \cite{Maciejewski2020WHAMR} &-  &- &12.0 \\
		& &Cascaded System \cite{Maciejewski2020WHAMR} &- &- &12.6 \\\cline{2-6}
		& \multirow{4}{*}{SE} &SpEx \cite{spex2020} &7.3s (avg) &13.2 &12.4 \\
		& &SpEx+ \cite{spex_plus2020} &7.3s (avg) &14.0 &13.5 \\\cline{3-6}
		& &\multirow{2}{*}{SpEx++ (2-stages)} &7.3s (avg) &14.5 &14.0 \\ 
		& & &60s &\textbf{14.8} &\textbf{14.3}  \cr
		\hline
		\hline
		\multirow{7}{*}{Reverb}&\multirow{3}{*}{BSS} &Conv-TasNet \cite{Maciejewski2020WHAMR} &-  &- &7.6\\
		& &BLSTM-TasNet \cite{Maciejewski2020WHAMR} &-  &- &8.9 \\
		& &Cascaded System \cite{Maciejewski2020WHAMR} &- &- &9.9 \\\cline{2-6}
		& \multirow{4}{*}{SE} &SpEx \cite{spex2020} &7.3s (avg) &8.7 &9.7 \\
		& &SpEx+ \cite{spex_plus2020} &7.3s (avg) &9.3 &10.6 \\\cline{3-6}
		& &\multirow{2}{*}{SpEx++ (2-stages)} &7.3s (avg) &9.7 &11.0 \\ 
		& & &60s &\textbf{10.0} &\textbf{11.4}  \cr
		\hline
		\hline
		\multirow{7}{*}{\shortstack{Noise\\+\\Reverb}}&\multirow{3}{*}{BSS} &Conv-TasNet \cite{Maciejewski2020WHAMR} &-  &- &8.3\\
		& &BLSTM-TasNet \cite{Maciejewski2020WHAMR} &-  &- &9.2 \\
		& &Cascaded System \cite{Maciejewski2020WHAMR} &- &- &10.1 \\\cline{2-6}
		& \multirow{4}{*}{SE} &SpEx \cite{spex2020} &7.3s (avg) &9.5 &10.3 \\
		& &SpEx+ \cite{spex_plus2020} &7.3s (avg) &10.0 &10.9 \\\cline{3-6}
		& &\multirow{2}{*}{SpEx++ (2-stages)} &7.3s (avg) &10.4 &11.4 \\ 
		& & &60s &\textbf{10.7} &\textbf{11.7}  \cr
		\hline
	\end{tabular} \label{tbl:cmp_whamr}
	\vspace{-10pt}
\end{table}

\vspace{-5pt}
\subsection{Results on noise-free condition}

We firstly compare  SpEx++ with previous baseline SpEx+ system on WSJ0-2mix in terms of SDRi, SI-SDRi and PESQ with both 7.5 seconds and 2 seconds reference speech.

From Table \ref{tbl:cmp_system_wsj0_2mix}, we conclude: 1) Multi-stage framework significantly outperforms single-stage framework. The improvements mainly come from the augmented frame-level speech embedding through a multi-stage pipeline. As an evidence, the 2-stage SpEx++ outperforms 1-stage counterpart substantially benefiting from the extra frame-level speaker embedding. However, the 3-stage SpEx++ doesn’t improve much further because the 3-stage and 2-stage solutions use similar speaker information overall (i.e., Utt + Fm). 
2) The signal fusion strategy in SpEx++ (1-stage) achieves 0.3dB performance gain over SpEx+ baseline in terms of SDRi and SI-SDRi. The results validate the effectiveness of the proposed signal fusion strategy. Besides, the learned weights for $\omega_1, \omega_2, \omega_3$ were 1.18, 0.32, and 0.11, respectively. This suggests that the decoded signal with higher temporal resolution contribute more. Note that we don't constrain that the sum of weights is equal to one.
3) The experiments with 2-second reference speech further show that SpEx++ system works well with much shorter reference speech, achieving 1.0dB SI-SDR improvement over SpEx+ baseline system.




\renewcommand{\arraystretch}{1.5}
\begin{table}[tp]
	
	\centering
	\fontsize{7.5}{5}\selectfont
	\caption{SDRi (dB) and SI-SDRi (dB) in an universal study of the
SpEx++ system on WHAMR!. The system is trained on 4 conditions and tested on each condition individually.} 
	\begin{tabular}{|c|c|c|c|c|c|}
		\hline
		Given & \multirow{2}{*}{Methods} &\multicolumn{2}{c|}{Conditions} &SDRi &SI-SDRi\\ \cline{3-4}
		Ref. &  &Training &Test &(dB) &(dB)\cr
		\hline
		\hline
		\multirow{6}{*}{\shortstack{7.3s\\(avg)}} &\multirow{3}{*}{SpEx \cite{spex2020}} &\multirow{3}{*}{4 conditions} &Noise &13.5 &12.7 \\
		& &  &Reverb &8.8 &9.8 \\
		& &  &Noise + Reverb &10.1 &10.8 \\\cline{2-6}
		& \multirow{3}{*}{\shortstack{SpEx++\\(2-stages)}} &\multirow{3}{*}{4 conditions} &Noise &14.6 &14.2\\
		& &  &Reverb &9.9 &11.2\\
		& &  &Noise + Reverb &10.7 &11.6\\
		\hline
		\hline
		\multirow{6}{*}{60s} &\multirow{3}{*}{SpEx \cite{spex2020}} &\multirow{3}{*}{4 conditions} &Noise &14.3 &13.8 \\
		& &  &Reverb &9.7 &10.8 \\
		& &  &Noise + Reverb &10.8 &11.7 \\ \cline{2-6}
		& \multirow{3}{*}{\shortstack{SpEx++\\(2-stages)}} &\multirow{3}{*}{4 conditions} &Noise &14.9 &14.6\\
		& &  &Reverb &10.3 &11.5\\
		& &  &Noise + Reverb &11.1 &12.0\\
		\hline
	\end{tabular} \label{tbl:cmp_universal}
	\vspace{-12pt}
\end{table}

\vspace{-3pt}
\subsection{Results on noisy conditions}

We further verify the robustness of SpEx++ system with other baselines under noisy conditions, as reported in Table \ref{tbl:cmp_wham} and Table \ref{tbl:cmp_whamr}. Compared with the results under noise-free condition in Table 1, we observe that the additive noise and reverberation in the mixture significantly degrades the performance of both blind speech separation and speaker extraction systems. This fact shows that extracting target speaker's voice from noisy mixture is a challenging task. Despite this, our SpEx++ system still shows effective performance under noisy conditions. Specifically, SpEx++ system achieves about 0.9dB and 0.5dB absolute improvement over SpEx+ baseline in terms of SI-SDRi on WHAM! and WHAMR!, respectively. 

We finally study an universal SpEx++ system that is trained on four mixture conditions (clean, noisy only, reverberation only, and noisy and reverberation). The performance of the universal system is further evaluated on each noisy mixture condition individually, as shown in Table \ref{tbl:cmp_universal}. Compared with the results in Table \ref{tbl:cmp_whamr} and Table \ref{tbl:cmp_universal}, we observe that the universal SpEx++ trained on four conditions achieves better performance than that trained on a single condition.

\vspace{-5pt}
\section{Conclusions}
\vspace{-5pt}
In this paper, we proposed a multi-stage time-domain speaker extraction called SpEx++, to alleviate the problem of insufficient reference materials of the target speaker. We took the advantages of multi-stage architecture to refine the speaker characteristics of target speaker using the extracted voice in previous stages. Experiments showed that the proposed frame-level sequential speech embedding serves as the second reference signal, and significantly improve speaker extraction without additional reference speech.



\vfill\pagebreak

\bibliographystyle{IEEEbib}
\bibliography{strings,refs}

\end{document}